\documentclass[apj]{emulateapj}
\usepackage{amsmath,amssymb,graphicx,natbib,longtable,bbding}
\usepackage[dvips]{color}

\newcommand{\etal}{et~al.\ }
\newcommand{\msun}{\mathcal{M}_\odot}
\newcommand{\mstellar}{\mathcal{M}_\star}
\newcommand{\sersic}{S\'ersic}

\slugcomment{Accepted for publication in the Astrophysical Journal}

\begin{document}

\title{The ACS Fornax Cluster Survey. X. Color Gradients of Globular Cluster
Systems in Early-Type Galaxies\altaffilmark{1}}

\author{Chengze Liu\altaffilmark{2},
Eric W. Peng\altaffilmark{2}, Andr\'es Jord\'an\altaffilmark{3,4},
Laura Ferrarese\altaffilmark{5}, John P. Blakeslee\altaffilmark{5},\\
Patrick C\^ot\'e\altaffilmark{5} and Simona Mei\altaffilmark{6,7}}

\altaffiltext{1}{Based on observations with the NASA/ESA Hubble
  Space Telescope, obtained at the Space Telescope Science Institute
  (STScI), which is operated by the Association of Universities for
  Research in Astronomy, Inc., under NASA contract \mbox{NAS
  5-26555}.}
\altaffiltext{2}{Department of Astronomy, Peking University, Beijing
  100871, China; czliu@pku.edu.cn, peng@bac.pku.edu.cn}
\altaffiltext{3}{Departmento de Astronom\'ia y Astrof\'isica,
  Pontificia Universidad Cat\'olica de Chile, Av.\ Vicu\~na Mackenna
  4860, 7820436 Macul, Santiago, Chile; ajordan@astro.puc.cl}
\altaffiltext{4}{Harvard-Smithsonian Center for Astrophysics, 60
  Garden St, Cambridge, MA 02138}
\altaffiltext{5}{Herzberg Institute of Astrophysics, Victoria, BC V9E
  2E7, Canada; laura.ferrarese@nrc-cnrc.gc.ca, John.Blakeslee@nrc-cnrc.gc.ca,
  patrick.cote@nrc-cnrc.gc.ca}
\altaffiltext{6}{University of Paris Denis Diderot, 75205 Paris
  Cedex 13, France}
\altaffiltext{7}{GEPI, Observatoire de Paris, Section de Meudon, 5
  Place J. Janssen, 92195 Meudon Cedex}

\begin{abstract}

We use the largest homogeneous sample of globular clusters (GCs),
drawn from the ACS Virgo Cluster Survey (ACSVCS) and ACS Fornax
Cluster Survey (ACSFCS), to investigate the color gradients of GC
systems in 76 early-type galaxies.  We find that most GC systems
possess an obvious negative gradient in ($g$--$z$) color with radius
(bluer outwards), which is consistent with previous work. For GC
systems displaying color bimodality, both metal-rich and metal-poor
GC subpopulations present shallower but significant color gradients
{\it on average}, and the mean color gradients of these two
subpopulations are of roughly equal strength. The field-of-view of
ACS mainly restricts us to measuring the inner gradients of the
studied GC systems. These gradients, however, can introduce an aperture bias
when measuring the mean colors of GC subpopulations from relatively
narrow central pointings. Inferred corrections to previous work imply
a reduced significance for the relation between the mean color of
metal-poor GCs and their host galaxy luminosity. The GC color
gradients also show a dependence with host galaxy mass where the
gradients are weakest at the ends of the mass spectrum---in massive
galaxies and dwarf galaxies---and strongest in galaxies of
intermediate mass, around a stellar mass of $\mstellar \approx
10^{10}\msun$. We also measure color gradients for field stars in the
host galaxies. We find that GC color gradients are systematically
steeper than field star color gradients, but the shape of the
gradient--mass relation is the same for both. If gradients are caused
by rapid dissipational collapse and weakened by merging, these color
gradients support a picture where the inner GC systems of most
intermediate-mass and massive galaxies formed early and rapidly with
the most massive galaxies having experienced greater merging. The
lack of strong gradients in the GC systems of dwarfs, which
probably have not experienced many recent major mergers, suggests
that low mass halos were inefficient at retaining and mixing metals
during the epoch of GC formation.

\end{abstract}

\keywords{galaxies: stellar content -- clusters: individual (Virgo,
Fornax) -- galaxies: elliptical and lenticular, cD -- galaxies: star
clusters -- globular clusters: general}

\section{Introduction}
\label{sec:intro}

Galactic radial gradients in stellar populations are a result of a
galaxy's star formation, chemical enrichment, and merging histories,
and thus can be an important discriminant of galaxy formation
scenarios.  Galaxies that form in a strong dissipative collapse are
expected to have steep gradients in metallicity, as the central
regions retain gas more effectively and form stars more efficiently.
Thus, in isolation, higher mass galaxies formed in this way are
expected to have steeper negative metallicity gradients due to their
deeper potential wells (e.g., \citealt{Chiosi2002_MNRAS_335_335,
Kawata2003_MNRAS_340_908}).  By contrast, in galaxies where  merging
is a dominant process, radial gradients are expected to weaken due to
radial mixing that occurs during mergers
\citep{White1980_MNRAS_191_1, Bekki1999_ApJ_513_108,
Kobayashi2004_MNRAS_347_740}. So if the most massive, quiescent
galaxies are the ones most shaped by major merging (e.g.,
\citealt{van_der_wel2009ApJ_706_120}), one would expect their
metallicity gradients to be relatively flat. Recently, however,
\citet{Pipino2010_MNRAS_407_1347} argue that shallow gradients in
massive galaxies can also result from lower star formation efficiency
and do not necessarily require extensive merging.

The existence of negative optical and near-infrared color gradients,
where the outer regions are bluer, have been well-established in
elliptical and disk galaxies (e.g., \citealt{Franx1989_AJ_98_538,
Peletier1990_AJ_100_1091, Michard2005_A+A_441_451,
Wu2005_ApJ_622_244, Liu2009_RAA_9_1119}), and have generally been
interpreted as gradients in metallicity, or sometimes age (e.g.,
\citealt{Kobayashi1999_ApJ_527_573, Kuntschner2006_MNRAS_369_497,
Rawle2008_MNRAS_389_1891}). In lower mass galaxies, however,
gradients appear to be shallower, nonexistent, or even positive.
This shows that gradient properties can be
a function of galaxy mass and perhaps reflects the greater
diversity in the star formation and evolutionary histories of
low-mass galaxies. Recent results with large samples of galaxies
show that while the most massive galaxies have shallow or flat color
gradients, gradients get increasingly negative toward lower stellar
mass until $\mstellar\sim 3\times10^{10} \msun$, at which point
gradients again become shallower and even positive
\citep{Spolaor2009_ApJ_691_138, Tortora2010_MNRAS_407_144}.  For
early-type galaxies in particular, this has been interpreted as an
intrinsic correlation between gradient and galaxy mass---more
negative at higher mass---modulated by dry merging at higher masses,
especially for brightest cluster galaxies
\citep{Roche2010_MNRAS_407_1231}.

Nearly all previous studies of stellar population gradients are of
the main stellar body (bulge or disk) of a galaxy.  Given the complex
star formation histories of galaxies, the effects of age and
metallicity are often difficult to disentangle, and require multiband
photometry and spectroscopy (e.g.,
\citealt{MacArthur2004_ApJS_152_175}).  Moreover, these studies say
little about the stellar halo, perhaps the oldest galactic component.
We thus approach the issue of population gradients using a unique
tool: globular clusters.

Globular clusters (GCs) are among the oldest stellar populations in
galaxies, and preserve information from the earliest epochs of star
formation. Population gradients in GC systems have not received much
attention, but one notable exception was the study of metallicity
gradients in the Milky Way GC system by
\citet{Searle1978_ApJ_225_357}.  They showed that although the inner
halo GCs had a negative gradient, the outer halo GCs had no gradient,
leading them to suggest that the outer halo was accreted from
dwarf-like fragments.

In both the Milky Way and nearby galaxies, GCs are found to be nearly
universally old, with ages greater than $\sim8$~Gyr (e.g.,
\citealt{Puzia2006_ApJ_648_383, Hansen2007_ApJ_671_380,
Mar'in-Franch2009_ApJ_694_1498, Woodley2010_ApJ_708_1335}). Although
in extragalactic systems we are mostly limited to broadband colors,
the lack of any significant age spread in GCs, and the fact that they
are generally simple stellar populations, allows us to interpret GC
colors as largely representative of metallicity.

The color distributions of GCs in massive galaxies are often bimodal,
and usually interpreted as two metallicity subpopulations
\citep[e.g., ][]{Gebhardt1999_AJ_118_1526, Larsen2001_AJ_121_2974,
Kundu2001_AJ_121_2950, Peng2006_ApJ_639_95} (although there is still
uncertainty in the transformation from color to metallicity, see
Yoon, Yi \& Lee 2006). Metal-rich (red) GCs are found to have a more
concentrated spatial distribution than the metal-poor (blue) GCs,
which results in the total mean color of GCs becoming gradually bluer
with projected radius \citep[e.g.,][] {Rhode2001_AJ_121_210,
Jord'an2004_AJ_127_24, Tamura2006_MNRAS_373_601}.

Many studies of massive galaxies have confirmed that GC systems taken
as a whole have negative color and metallicity gradients \citep{
Geisler1996_AJ_111_1529, Rhode2001_AJ_121_210, Jord'an2004_AJ_127_24,
Cantiello2007_ApJ_668_209}. The conventional wisdom, however, has
been that individual metal-rich or metal-poor GC subpopulations have
no color or metallicity gradients \citep{Lee1998_AJ_115_947,
Rhode2001_AJ_121_210}. Additional studies of individual galaxies,
however, have shown that GC subpopulations in M49, M87, NGC~1427, and
NGC~1399, and nearby brightest cluster galaxies do have a slightly
negative color gradients \citep{Geisler1996_AJ_111_1529,
Forte2001_AJ_121_1992, Bassino2006_A+A_451_789,
Harris2009_ApJ_703_939, Harris2009_ApJ_699_254}. Furthermore, very
little is known about color gradients in the GC systems of dwarf
galaxies, whose systems are dominated by metal-poor GCs.  Similar to
population gradient studies of the main bodies of galaxies,
investigating the color or metallicity gradients of GC systems across
a range of galaxy mass can provide direct constraints on the
formation of GC systems and the merging history of their host
galaxies.

In this paper, we present the results from the first homogeneous
study of color gradients in the GC systems of early-type galaxies.
The ACS Virgo Cluster Survey (ACSVCS,
\citealt{Cot'e2004_ApJS_153_223}) and ACS Fornax Cluster Survey
(ACSFCS, \citealt{Jord'an2007_ApJS_169_213}) observed 100 galaxies in
the Virgo Cluster and 43 galaxies in the Fornax Cluster using the
Hubble Space Telescope Advanced Camera for Surveys (HST/ACS). All 143
objects are early-type galaxies and range in mass from dwarf to giant
galaxies. One of the main goals of the surveys is the investigation
of extragalactic GC systems, and previous studies have examined their
color distributions \citep{Peng2006_ApJ_639_95}, size distributions
\citep{Jord'an2005_ApJ_634_1002, Masters2010_ApJ_715_1419},
luminosity functions \citep{Jord'an2006_ApJ_651_25,
Jord'an2007_ApJS_171_101, Villegas2010_ApJ_717_603}, formation
efficiencies \citep{Peng2008_ApJ_681_197}, and color-magnitude
relations \citep{Mieske2006_ApJ_653_193, Mieske2010_ApJ_710_1672}.
Likewise, the surface photometry of the galaxies themselves have also
been studied in detail \citep{Ferrarese2006_ApJS_164_334,
Cot'e2007_ApJ_671_1456}, allowing us to perform a homogeneous
comparison of the color gradients in the field stars with those in
the GC systems. Another advantage of this sample is that distances to
most galaxies have been determined using the method of surface
brightness fluctuations \citep{Mei2007_ApJ_655_144,
Blakeslee2009_ApJ_694_556}. Using this large and homogenous sample of
extragalactic GCs \citep{Jord'an2009_ApJS_180_54}, we measure the
color gradients of GC systems in the targeted galaxies within the
field of view (FOV) of the ACS camera, except for four galaxies where
we use multiple ACS fields. The high resolution and quality of the
HST images allow us to measure the gradients of GC systems in dwarf
galaxies as well as in individual GC subpopulations for systems
showing color bimodality.

This paper is organized as follows: In Section~2, we give a
description of the GC selection and data analysis. The results and
discussion are presented in Sections~3 and 4, respectively. Finally,
we conclude in Section~5.

\section{Sample and Data Analysis}
\label{sec:sample}

\subsection{Galaxy Sample and GC Selection}
\label{sec:galaxy sample}

The data used in this work are drawn from the ACSVCS and ACSFCS,
which obtained deep, high-resolution images for 143 early-type
galaxies in the F475W ($\approx$ SDSS $g$) and F850LP ($\approx$ SDSS
$z$) filters using HST/ACS. These galaxies were selected by
morphology (E, S0, dE and dS0) and cover a range in luminosity,
$-22<M_B<-15$ (see \citealt{Cot'e2004_ApJS_153_223} and
\citealt{Jord'an2007_ApJS_169_213} for details).

Details about the selection of over 12,000 GC candidates in the 100
early-type Virgo galaxies are described in
\citet{Jord'an2004_ApJS_154_509, Jord'an2009_ApJS_180_54}.  Briefly,
after selecting preliminary GC candidates using magnitude,
ellipticity, and a broad color cut, all the candidates are fit with a
PSF-convolved King model using the KINGPHOT code. The probability of
an object being a GC (the $p_{\rm GC}$ parameter) is determined in
the plane of magnitude and half-light radius with comparison to a
number of control fields. GCs in the 43 Fornax galaxies were selected
using the same method. Although previous studies have used a
criterion of $p_{\rm GC}>0.5$ for GCs, in this work, we select GCs
with $p_{\rm GC}>0.95$. The reason we choose this stringent criterion
is that for the outer regions of dwarf galaxies, the contamination
from background galaxies is the limiting factor.  Such a strict
selection causes us to lose fainter GCs (affecting our completeness),
but increases our efficiency.  This stricter cut in the $p_{\rm GC}$
parameter actually introduces a varying completeness with galaxy
mass---essentially, galaxies with more GCs have fainter limits---but
we have run Monte Carlo simulations to show that our results do not
change if we choose a simple magnitude limit for all galaxies. Our
more detailed but still rigorous approach to selection allows us to
optimize signal-to-noise, especially for bimodal color distributions
where we are splitting the sample in two.

Contamination by compact background galaxies is one of our main
problems.  To estimate the contamination from foreground and
background, we used 16 control fields at high latitude (Table~1 of
\citealt{Peng2006_ApJ_639_95}). As described in detail by
\citet{Peng2006_ApJ_639_95} and \citet{Jord'an2009_ApJS_180_54}, the
expected contamination was estimated for each target galaxy. We have
checked that the contamination is negligible if we select GCs with GC
probability $p_{\rm GC}>0.95$, averaging $\sim1$ object per ACS
field.

\subsection{GC Subpopulations}
\label{sec:GC subpopulations}

Previous studies show that most massive galaxies have bimodal GC
color distributions (e.g. \citealt{Gebhardt1999_AJ_118_1526,
Larsen2001_AJ_121_2974, Kundu2001_AJ_121_2950,
Spitler2008_MNRAS_389_1150}). \citet{Peng2006_ApJ_639_95} presented
the color distributions for GC systems in 100 ACSVCS galaxies.
Following their work, we use Kaye's Mixture Model (KMM;
\citealt{McLachlan1988_mmia.book__, Ashman1994_AJ_108_2348}) to
decompose the data into two Gaussian distributions with the same
$\sigma$ (homoscedastic).  We choose the homoscedastic case because
it is more stable for small samples.  In practice, for galaxies with
large numbers of GCs, allowing $\sigma$ to vary has no effect on
these results.  For each GC system, we determine which GCs are
members of the blue and red GC subpopulations and the '$p$-value'
(not to be confused with $p_{\rm GC}$) for the bimodal model. We
consider the GC color distribution to be bimodal if the '$p$-value'
is less than 0.05. A total of 40 galaxies meet this criterion.
Membership in the red or blue subpopulation is determined by the
membership probabilities output by KMM, and corresponds to the
``dip'' in the color distribution.   If the $p$-value is not less
than 0.05, the galaxy is deemed to have only one GC population.

We show two galaxies as examples in Figure~\ref{fig:cp}. The right
panels of Figure \ref{fig:cp} show the color distributions for GCs in
FCC~47 (NGC~1336, panel $c$) and FCC 153 (IC~335) panel $d$). The GC
color distribution of FCC~47 displays two peaks, while the GCs in
FCC~153 has just one peak in color. The blue and red curves in panel
$c$ are Gaussian fits to the blue and red GCs determined by KMM. In
panel $d$, the black curve shows the best fitting of color
distribution of whole GC systems using a single Gaussian function.

\subsection{Calculating Radial Gradients}
\label{sec:gradients calculation}

\begin{figure}[]
  \includegraphics[width=0.48\textwidth]{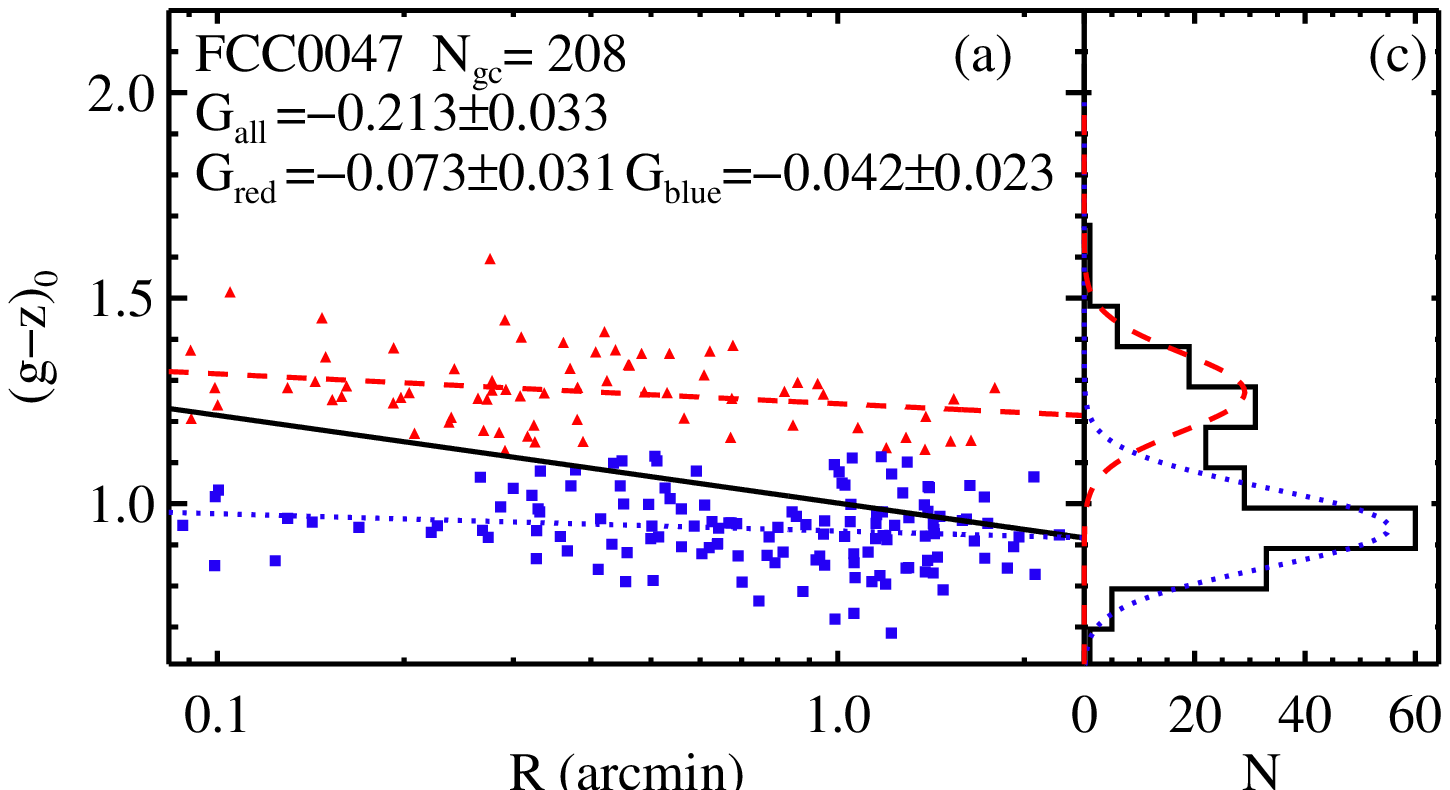}
  \includegraphics[width=0.48\textwidth]{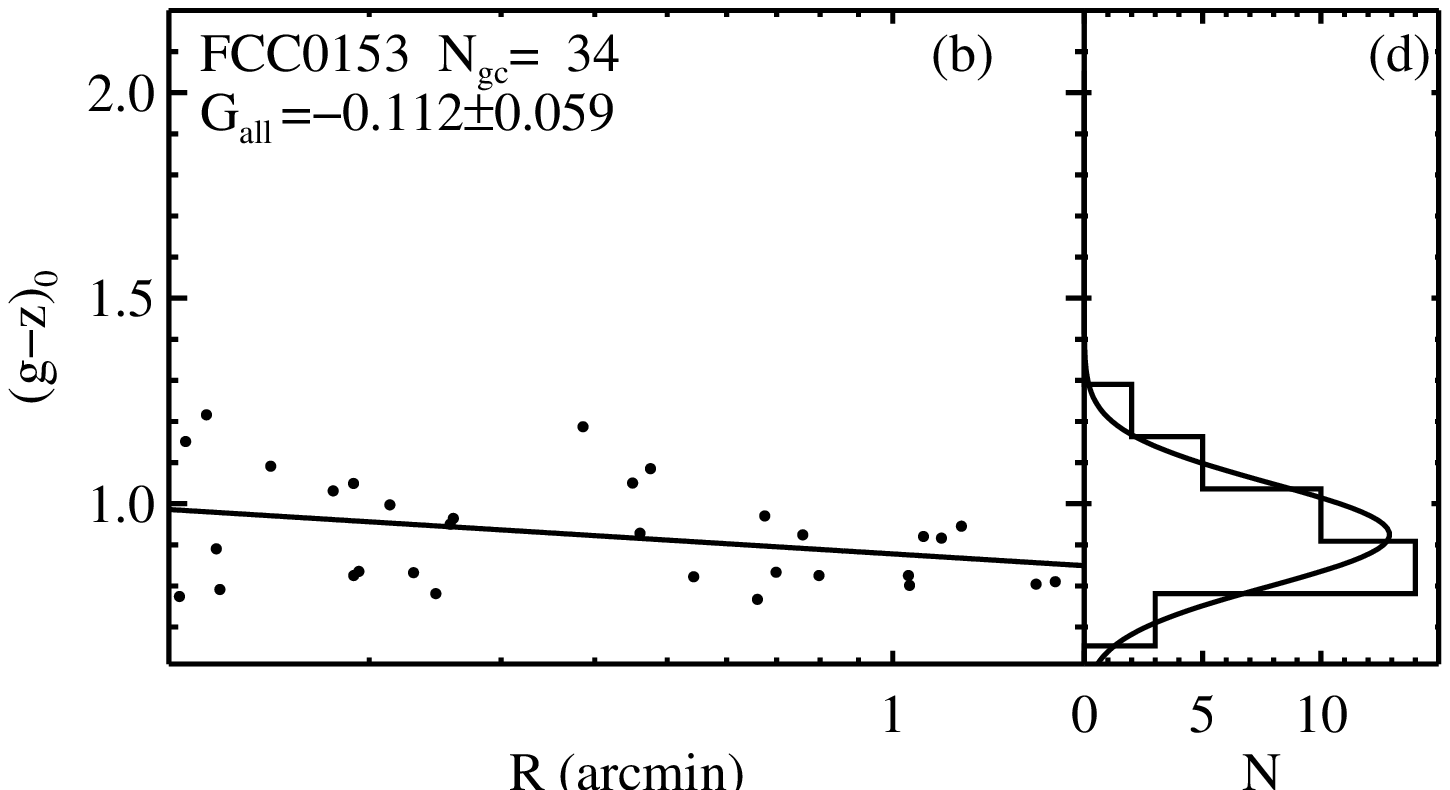}\\
  \caption{Color profiles (left panels) and color distributions
  (right panels) of GC systems in two sample galaxies.
  a) GC system color profile of FCC~47 (NGC~1336), a GC system with a
  bimodal color distribution. Each small dot denotes a
  GC color coded for the blue and red subpopulation. The blue, red
  and black lines are the best linear fit of metal-poor, metal-rich
  and total GC populations.
  b) GC system color profile of unimodal galaxy FCC~153 (IC~335).
  c) GC color distribution of GCs in FCC~47. Blue and red curves
  represent the Gaussian fitting of blue and red GCs determined by KMM.
  d) GC color distribution for FCC~153, black curve represents
  the Gaussian fitting of all GCs.}\label{fig:cp}
\end{figure}

The radial gradients are calculated by a linear least-squares fit
between the GC color or metallicity and the logarithm of the radius,
defined as:
\begin{eqnarray}
   G_{g-z}  &=& \frac{\Delta (g-z)}{\Delta \log R} \label{eqn:cgrad}\\
   G_{[Fe/H]} &=& \frac{\Delta [Fe/H]}{\Delta \log R }
\end{eqnarray}
In other words, we measure the change in color or metallicity per dex
in radius. The color gradient errors are one-sigma errors and come
from linear regression. To ensure adequate signal-to-noise, we
restrict ourselves to the 78 galaxies with at least 20 GCs that meet
our selection criteria ($p_{\rm GC}>0.95$). We subsequently eliminate
VCC~1938 from our sample because of its close projected separation
from the dwarf elliptical, VCC~1941. We also remove the S0/a
transition galaxy, FCC167, due to uncertainties in measuring its
stellar mass. This leaves us with a sample of 76 galaxies. If the
galaxy has a bimodal GC color distribution, we calculate both the
color gradient of whole GC system and the color gradient of each GC
subpopulation.  We divide the GCs in to red and blue using a simple
color cut determined by the KMM probabilities. We tried various
approaches, including running KMM as a function of radius, and
allowing the dividing color to vary with radius using an iterative
fitting process.  In the former case, the number of GCs limited the
effectiveness to only a handful of galaxies, and in the latter case
our results did not change in a significant way so we ultimately
chose to use the simplest method.  Given the FOV of ACS
($3\farcm4\times3\farcm4$), the maximum possible outer radius of our
gradient measurements is $\sim2\farcm4$, which corresponds to 11.5
and 14.0~kpc at the distances of the Virgo and Fornax clusters, respectively.

The left panels of Figure \ref{fig:cp} show color profiles of GC
systems in FCC~47 (panel $a$) and FCC 153 (panel $b$). We can see
from the figure that both red and blue GC systems in FCC~47 show
negative color gradients with color becoming gradually bluer from the
center to the outskirts of the galaxy. The gradients of the whole GC
system is steeper than that of each subpopulation due to the
dominance of blue GCs at large radii. The GC system in the unimodal
galaxy FCC~153 also shows a shallower but significant negative color
gradient.

For the two most luminous giant galaxies in the Virgo cluster: M49
(VCC~1226) and M87 (VCC~1316), the gradients are extended by
including the GCs in nearby ACS fields. Because some targeted
galaxies were located in the halos of the giants, and their own GC
systems appear to be entirely stripped (see
\citealt{Peng2008_ApJ_681_197} for details), we consider the GCs in
these fields as part of the GC systems of the giant galaxies. For
M49, we use VCC~1199 and 1192, extending our study to a radius of
$4\farcm5$ (22~kpc). For M87, we use VCC~1327, 1297, 1279,
1185 and 1250, which extends our study to a radius of
$21\farcm3$ (102~kpc).  There are two similar cases in the Fornax cluster.
FCC~202 is near FCC~213 ($4\farcm6$, 27~kpc)  and FCC~143 is
near FCC~147  ($4\farcm8$, 28~kpc).

\section{Results}
\label{sec:results}

\subsection{Color Gradients}
\label{sec:color gradients}

\begin{figure}[]
  \includegraphics[width=0.48\textwidth]{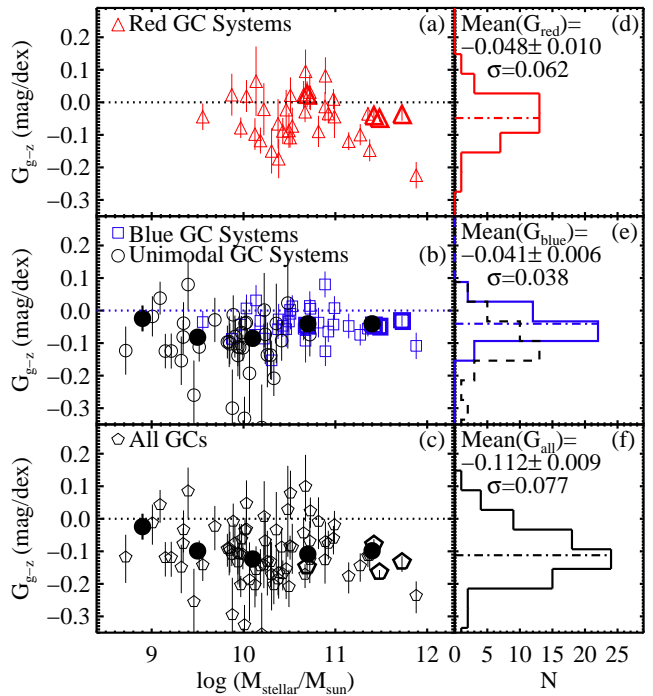}\\
  \caption{Color gradients as a function of galaxy stellar mass (left
    panels) and gradient distributions (right panels).
  From top to bottom: red GC systems, blue and unimodal GC systems,
  and all GCs. The dotted lines in panel $a$, $b$, and $c$ denote
  the zero gradients. In panel $d$, $e$ and $f$,
  the dot-dashed lines describe the mean $G_{g-z}$ of red, blue and
  whole GC systems respectively. The dashed histogram in panel $e$
  is the distribution of color gradients in unimodal galaxies.
  Big filled circles in panel $b$
  and $c$ denote the mean gradients in given mass bins.
  The larger and thicker open symbols at the high mass ends of panels $a$ $b$ and
  $c$, denote the four galaxies whose profiles were extended by the
  use of neighboring ACS fields: M49, M87, FCC~213, and FCC~147 (see Section
  \ref{sec:gradients calculation}).\label{fig:cg_mass}}
\end{figure}

Figure \ref{fig:cg_mass} presents results for all 76 galaxies,
showing the color gradient distribution (right panels), and the
strength of the color gradients as a function of galaxy stellar mass
(left panels).  The stellar masses for the ACSVCS galaxies were taken
from \citet{Peng2008_ApJ_681_197}, and the masses for the ACSFCS
galaxies were calculated in the same way as described in that paper
using $g$--$z$ photometry from the ACS images (Ferrarese et al., in
prep) and $J$--$K$ colors from the Two Micron All Sky Survey (2MASS,
\citealt{Skrutskie2006_AJ_131_1163}). From top to bottom, this figure
shows the color gradients of red GC populations, blue GC and unimodal
populations, and whole GC systems, respectively. We combine the blue
GCs and unimodal populations on the same plot because unimodal
populations for low mass galaxies consist nearly entirely of blue
GCs, and are likely the low mass extension of the blue GCs in more
massive galaxies (see \citealt{Peng2006_ApJ_639_95}). Consistent with
previous studies (e.g. \citealt{Geisler1996_AJ_111_1529,
Rhode2001_AJ_121_210, Jord'an2004_AJ_127_24,
Tamura2006_MNRAS_373_601, Harris2009_ApJ_699_254}), we find that the
whole GC systems of most giant early-type galaxies have negative
color gradients. We also find that not only giant galaxies, but also
most intermediate- and low-mass galaxies show shallow but significant
color gradients in their GC systems.

As described in Section~\ref{sec:GC subpopulations} and
\ref{sec:gradients calculation}, we calculate the color gradients of
individual red and blue GC systems respectively if GCs display
bimodal color distribution. Figure \ref{fig:cg_mass} shows that although
the subpopulation gradients for individual galaxies are often not by
themselves very significant, both red and blue GC systems have
significant shallow negative color gradients when we combine data from
many galaxies.  The red and blue GC gradients have mean values
equaling $-0.048 \pm
0.010$ and $-0.041 \pm 0.006$ respectively. The errors in the color
gradients of individual GC systems are taken into account when
calculating the mean color gradients and their errors. Color gradients of red GC
populations are slightly steeper than that of blue GC systems and
seem to show more scatter with dispersions $\sigma_{\rm{red}}=0.062$
and $\sigma_{\rm{blue}}=0.038$. Furthermore, both red and blue GC
systems individually have much shallower color gradients than that of
whole GC systems ($-0.112 \pm 0.009$) with dispersion
$\sigma_{\rm{all}}=0.077$.

Table \ref{tab:cg_VCC} and \ref{tab:cg_FCC} lists the color gradients
of blue, red and whole GC systems for our sample galaxies. We only
show the results for the 76 galaxies with more than 20 GCs that meet
our selection.

In panels $b$ and $c$ of Figure \ref{fig:cg_mass}, big filled circles
display the mean color gradients in given mass bins with bin widths
of 0.6~dex. For galaxies with $\mstellar\lesssim 10^{10}\msun$, there
appears to be a weak correlation between color gradients and galaxy
mass, with color gradients tending to be shallower for dwarf
galaxies. But for the higher mass galaxies, the trend is flattened,
even reversed. In this figure, Virgo and Fornax galaxies are plotted
together. We do not find a significant difference in behavior between
galaxies in the different clusters.

\subsection{Metallicity Gradients}

Since most GCs are old, single stellar populations, trends in their
integrated color are generally equated with trends in metallicity.
Recent studies have found a non-linear but monotonic relation between
metallicity and color of GCs (e.g., \citealt{Harris2002_AJ_123_3108,
Peng2006_ApJ_639_95, Blakeslee2010_ApJ_710_51}), with broadband color
less sensitive at lower metallicity. \citet{Blakeslee2010_ApJ_710_51}
fit the color-metallicity relation from the data shown in
(\citealt{Peng2006_ApJ_639_95}) using a quartic polynomial (their
Equation~1). Although the conversion from color to metallicity is
still uncertain and contains considerable scatter, we can use the
\citet{Blakeslee2010_ApJ_710_51} relation to derive a radial
metallicity profile for each GC system. After conversion, the
metallicity distribution of GC systems in many galaxies are not
bimodal (see \citealt{Yoon2006_Sci_311_1129,
Cantiello2007_ApJ_669_982, Blakeslee2010_ApJ_710_51}), but
interpreting this is beyond the scope of this paper (see also
\citealt{Spitler2008_MNRAS_389_1150}). In this work, we only use this
relation to calculate the mean metallicity gradients of the entire GC
system of each galaxy.

\begin{figure}[!t]
  \includegraphics[width=0.48\textwidth]{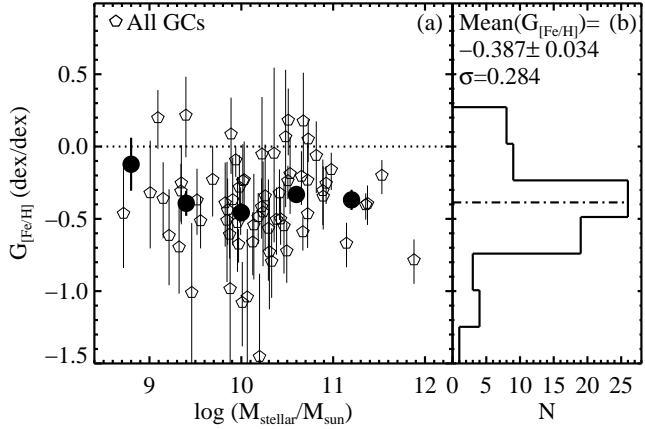}\\
  \caption{The same as panel $c$ and $f$ of Figure \ref{fig:cg_mass},
  but with color gradients converted to metallicity for whole GC
  systems.}\label{fig:histogram_metal}
\end{figure}

Figure \ref{fig:histogram_metal} presents the distribution of
metallicity gradients of all GCs and the gradient--mass relation.
Similar to the color gradients, the GC systems of most galaxies have
shallow but significant metallicity gradients with a mean value of
$-0.387 \pm 0.034$ with dispersion $\sigma_{[Fe/H]}=0.284$. The
metallicity gradient--mass relation is also similar to the color
gradient--mass relation shown in Figure \ref{fig:cg_mass}.

{\centering
\begin{deluxetable*}{lrrrrrrr}
 \tablewidth{0pt}
 \tabletypesize{\scriptsize}
 \tablecaption{Color and Metallicity gradients of GC systems of ACSVCS galaxies
 \label{tab:cg_VCC}}
 \tablehead{
  \colhead{Name} &
  \colhead{$N_{\textrm{GC}}$} &
  \colhead{$\log(\mstellar/\msun)$} &
  \colhead{$G_{\rm{\textrm{blue}}}$} &
  \colhead{$G_{\textrm{red}}$} &
  \colhead{$G_{\textrm{all}}$} &
  \colhead{$G_{\textrm{[Fe/H]}}$} &
  \colhead{$G_{\textrm{gal}}$}  \\
  \colhead{(1)} &
  \colhead{(2)} &
  \colhead{(3)} &
  \colhead{(4)} &
  \colhead{(5)} &
  \colhead{(6)} &
  \colhead{(7)} &
  \colhead{(8)}}
 \startdata
 VCC 1226 & $  950$ & $ 11.73$ & $-0.032\pm0.017$ & $-0.037\pm0.019$ & $-0.139\pm0.024$ & $-0.493\pm0.080$ & $-0.058\pm0.003$\\
 VCC 1316 & $ 2002$ & $ 11.48$ & $-0.049\pm0.008$ & $-0.049\pm0.011$ & $-0.171\pm0.013$ & $-0.570\pm0.044$ & $-0.015\pm0.002$\\
 VCC 1978 & $  709$ & $ 11.53$ & $-0.027\pm0.027$ & $-0.021\pm0.023$ & $-0.070\pm0.036$ & $-0.214\pm0.117$ & $-0.026\pm0.002$\\
 VCC  881 & $  256$ & $ 11.46$ & $-0.010\pm0.036$ & $-0.027\pm0.044$ & $-0.011\pm0.049$ & $-0.065\pm0.180$ & $ 9.999\pm0.000$\\
 VCC  798 & $  258$ & $ 11.27$ & $-0.075\pm0.033$ & $-0.099\pm0.033$ & $-0.150\pm0.044$ & $-0.543\pm0.145$ & $ 0.104\pm0.006$\\
 VCC  763 & $  408$ & $ 11.37$ & $-0.048\pm0.025$ & $-0.145\pm0.034$ & $-0.117\pm0.040$ & $-0.404\pm0.137$ & $-0.042\pm0.003$\\
 VCC  731 & $  772$ & $ 11.35$ & $-0.056\pm0.024$ & $-0.033\pm0.021$ & $-0.118\pm0.030$ & $-0.408\pm0.091$ & $-0.104\pm0.002$\\
 VCC 1535 & $  163$ & $ 10.89$ & $ 0.079\pm0.040$ & $-0.013\pm0.045$ & $-0.134\pm0.068$ & $-0.317\pm0.227$ & $ 0.133\pm0.154$\\
 VCC 1903 & $  244$ & $ 10.92$ & $-0.065\pm0.032$ & $-0.035\pm0.030$ & $-0.082\pm0.044$ & $-0.281\pm0.132$ & $-0.102\pm0.004$\\
 VCC 1632 & $  355$ & $ 10.98$ & $-0.045\pm0.030$ & $ 0.012\pm0.031$ & $-0.064\pm0.044$ & $-0.182\pm0.128$ & $-0.056\pm0.004$\\
 VCC 1231 & $  199$ & $ 10.73$ & $ 0.006\pm0.030$ & $ 0.017\pm0.037$ & $ 0.019\pm0.047$ & $ 0.035\pm0.149$ & $-0.056\pm0.003$\\
 VCC 2095 & $   75$ & $ 10.72$ &          \nodata &          \nodata & $-0.072\pm0.067$ & $-0.256\pm0.272$ & $-0.116\pm0.002$\\
 VCC 1154 & $  132$ & $ 10.89$ & $-0.125\pm0.045$ & $ 0.082\pm0.057$ & $-0.074\pm0.067$ & $-0.363\pm0.218$ & $-0.019\pm0.002$\\
 VCC 1062 & $  129$ & $ 10.72$ & $ 0.013\pm0.045$ & $ 0.032\pm0.053$ & $-0.130\pm0.067$ & $-0.473\pm0.224$ & $-0.086\pm0.002$\\
 VCC 2092 & $   52$ & $ 10.68$ & $-0.093\pm0.067$ & $ 0.095\pm0.067$ & $ 0.097\pm0.103$ & $ 0.163\pm0.345$ & $-0.132\pm0.003$\\
 VCC  369 & $  128$ & $ 10.51$ & $-0.002\pm0.048$ & $ 0.027\pm0.055$ & $ 0.073\pm0.075$ & $ 0.170\pm0.230$ & $-0.078\pm0.003$\\
 VCC  759 & $  112$ & $ 10.65$ & $-0.056\pm0.035$ & $ 0.023\pm0.042$ & $-0.048\pm0.052$ & $-0.221\pm0.187$ & $-0.060\pm0.006$\\
 VCC 1692 & $   93$ & $ 10.53$ & $ 0.011\pm0.035$ & $-0.073\pm0.057$ & $-0.094\pm0.077$ & $-0.205\pm0.267$ & $-0.077\pm0.006$\\
 VCC 1030 & $  118$ & $ 10.12$ & $-0.075\pm0.037$ & $-0.094\pm0.048$ & $-0.209\pm0.064$ & $-0.669\pm0.217$ & $-0.081\pm0.009$\\
 VCC 2000 & $  148$ & $ 10.38$ & $-0.057\pm0.030$ & $-0.072\pm0.074$ & $-0.188\pm0.045$ & $-0.684\pm0.171$ & $-0.092\pm0.002$\\
 VCC  685 & $  125$ & $ 10.49$ & $-0.035\pm0.042$ & $-0.087\pm0.052$ & $-0.210\pm0.058$ & $-0.734\pm0.208$ & $-0.026\pm0.008$\\
 VCC 1664 & $  104$ & $ 10.42$ &          \nodata &          \nodata & $-0.095\pm0.062$ & $-0.328\pm0.175$ & $-0.118\pm0.003$\\
 VCC  654 & $   23$ & $ 10.36$ &          \nodata &          \nodata & $-0.036\pm0.121$ & $-0.090\pm0.600$ & $-0.056\pm0.005$\\
 VCC  944 & $   62$ & $ 10.46$ & $-0.060\pm0.055$ & $-0.023\pm0.061$ & $-0.162\pm0.088$ & $-0.560\pm0.319$ & $-0.075\pm0.002$\\
 VCC 1720 & $   42$ & $ 10.31$ & $-0.154\pm0.060$ & $-0.149\pm0.069$ & $-0.178\pm0.105$ & $-0.733\pm0.388$ & $-0.118\pm0.002$\\
 VCC  355 & $   29$ & $ 10.20$ &          \nodata &          \nodata & $-0.345\pm0.131$ & $-1.455\pm0.586$ & $-0.083\pm0.005$\\
 VCC 1619 & $   44$ & $ 10.24$ &          \nodata &          \nodata & $-0.075\pm0.091$ & $-0.501\pm0.367$ & $-0.026\pm0.001$\\
 VCC 1883 & $   43$ & $ 10.22$ & $-0.079\pm0.046$ & $-0.021\pm0.080$ & $-0.144\pm0.067$ & $-0.452\pm0.207$ & $-0.018\pm0.004$\\
 VCC 1242 & $   78$ & $ 10.18$ & $-0.029\pm0.036$ & $-0.117\pm0.040$ & $-0.147\pm0.057$ & $-0.490\pm0.177$ & $-0.062\pm0.002$\\
 VCC  784 & $   43$ & $ 10.23$ &          \nodata &          \nodata & $-0.001\pm0.114$ & $-0.082\pm0.406$ & $-0.074\pm0.003$\\
 VCC 1537 & $   25$ & $ 10.01$ &          \nodata &          \nodata & $-0.342\pm0.065$ & $-1.068\pm0.294$ & $-0.058\pm0.003$\\
 VCC  778 & $   43$ & $ 10.26$ &          \nodata &          \nodata & $-0.137\pm0.099$ & $-0.387\pm0.382$ & $-0.064\pm0.003$\\
 VCC 1321 & $   22$ & $  9.84$ &          \nodata &          \nodata & $-0.085\pm0.078$ & $-0.630\pm0.426$ & $-0.084\pm0.004$\\
 VCC  828 & $   48$ & $ 10.14$ & $ 0.032\pm0.048$ & $ 0.065\pm0.107$ & $-0.159\pm0.077$ & $-0.553\pm0.365$ & $-0.038\pm0.002$\\
 VCC 1630 & $   29$ & $ 10.06$ &          \nodata &          \nodata & $-0.199\pm0.118$ & $-1.005\pm0.479$ & $-0.060\pm0.002$\\
 VCC 1146 & $   53$ & $  9.94$ &          \nodata &          \nodata & $-0.139\pm0.062$ & $-0.105\pm0.105$ & $-0.081\pm0.003$\\
 VCC 1025 & $   58$ & $ 10.33$ &          \nodata &          \nodata & $-0.210\pm0.055$ & $-0.801\pm0.239$ & $-0.113\pm0.003$\\
 VCC 1303 & $   37$ & $ 10.02$ &          \nodata &          \nodata & $-0.038\pm0.042$ & $-0.271\pm0.262$ & $-0.113\pm0.003$\\
 VCC 1913 & $   36$ & $ 10.03$ &          \nodata &          \nodata & $-0.037\pm0.070$ & $-0.235\pm0.275$ & $-0.079\pm0.004$\\
 VCC 1125 & $   39$ & $  9.91$ &          \nodata &          \nodata & $-0.080\pm0.064$ & $-0.382\pm0.302$ & $ 0.042\pm0.004$\\
 VCC 1475 & $   52$ & $  9.89$ &          \nodata &          \nodata & $-0.013\pm0.059$ & $ 0.067\pm0.266$ & $-0.044\pm0.004$\\
 VCC 1178 & $   58$ & $  9.85$ &          \nodata &          \nodata & $-0.102\pm0.063$ & $-0.450\pm0.228$ & $-0.005\pm0.004$\\
 VCC 1283 & $   36$ & $  9.96$ &          \nodata &          \nodata & $-0.113\pm0.099$ & $-0.533\pm0.324$ & $-0.047\pm0.003$\\
 VCC 1261 & $   22$ & $  9.69$ &          \nodata &          \nodata & $-0.033\pm0.070$ & $-0.202\pm0.235$ & $-0.007\pm0.004$\\
 VCC  698 & $   83$ & $  9.98$ &          \nodata &          \nodata & $-0.068\pm0.051$ & $-0.294\pm0.235$ & $ 0.018\pm0.006$\\
 VCC 1910 & $   34$ & $  9.32$ &          \nodata &          \nodata & $-0.158\pm0.074$ & $-0.703\pm0.315$ & $-0.017\pm0.005$\\
 VCC  856 & $   24$ & $  9.35$ &          \nodata &          \nodata & $-0.038\pm0.065$ & $-0.173\pm0.274$ & $ 0.001\pm0.006$\\
 VCC 1087 & $   43$ & $  9.52$ &          \nodata &          \nodata & $-0.110\pm0.045$ & $-0.383\pm0.228$ & $-0.057\pm0.010$\\
 VCC 1861 & $   28$ & $  9.46$ &          \nodata &          \nodata & $-0.253\pm0.107$ & $-0.974\pm0.479$ & $-0.009\pm0.007$\\
 VCC 1431 & $   40$ & $  9.34$ &          \nodata &          \nodata & $-0.078\pm0.054$ & $-0.316\pm0.196$ & $ 0.065\pm0.012$\\
 VCC 1528 & $   28$ & $  9.21$ &          \nodata &          \nodata & $-0.124\pm0.055$ & $-0.616\pm0.333$ & $-0.046\pm0.006$\\
 VCC 2019 & $   20$ & $  9.01$ &          \nodata &          \nodata & $-0.021\pm0.062$ & $-0.364\pm0.380$ & $-0.084\pm0.008$\\
 VCC 1545 & $   27$ & $  9.15$ &          \nodata &          \nodata & $-0.122\pm0.053$ & $-0.341\pm0.266$ & $-0.143\pm0.005$\\
 VCC 1407 & $   22$ & $  9.09$ &          \nodata &          \nodata & $ 0.040\pm0.050$ & $ 0.178\pm0.207$ & $ 0.011\pm0.004$\\
 VCC 1539 & $   24$ & $  8.72$ &          \nodata &          \nodata & $-0.079\pm0.073$ & $-0.363\pm0.359$ & $ 0.113\pm0.019$\\
 \enddata
 \tablenotetext{1}{The name of galaxies}
 \tablenotetext{2}{Total number of GCs those $p>0.95$}
 \tablenotetext{3}{Logarithm of stellar mass (in unit of $\msun$)}
 \tablenotetext{4}{Color gradient of red GCs with error, if bimodal}
 \tablenotetext{5}{Color gradient of blue GCs with error, if bimodal}
 \tablenotetext{6}{Color gradient of all GCs with error}
 \tablenotetext{7}{Metallicity gradient of all GCs with error}
 \tablenotetext{8}{Color gradient of galaxy with error}
 \tablenotetext{Note}{Due to the FOV of ACS, the outer boundaries
                      of our measurements of gradients in most
                      galaxies are about 2--3 arcmin.}
\end{deluxetable*}
}

{\centering
\begin{deluxetable*}{lrrrrrrr}
 \tablewidth{0pt}
 \tabletypesize{\scriptsize}
 \tablecaption{Color and Metallicity gradients of GC systems of
 ACSFCS galaxies
 \label{tab:cg_FCC}}
 \tablehead{
  \colhead{Name} &
  \colhead{$N_{\textrm{GC}}$} &
  \colhead{$\log(\mstellar/\msun)$} &
  \colhead{$G_{\rm{\textrm{blue}}}$} &
  \colhead{$G_{\textrm{red}}$} &
  \colhead{$G_{\textrm{all}}$} &
  \colhead{$G_{\textrm{[Fe/H]}}$} &
  \colhead{$G_{\textrm{gal}}$}  \\
  \colhead{(1)} &
  \colhead{(2)} &
  \colhead{(3)} &
  \colhead{(4)} &
  \colhead{(5)} &
  \colhead{(6)} &
  \colhead{(7)} &
  \colhead{(8)}}
 \startdata
 FCC   21 & $  231$ & $ 11.88$ & $-0.109\pm0.040$ & $-0.224\pm0.041$ & $-0.241\pm0.049$ & $-0.796\pm0.154$ & $-0.074\pm0.010$\\
 FCC  213 & $ 1067$ & $ 11.42$ & $-0.046\pm0.019$ & $-0.041\pm0.019$ & $-0.086\pm0.023$ & $-0.304\pm0.070$ & $-0.015\pm0.002$\\
 FCC  219 & $  297$ & $ 11.14$ & $-0.047\pm0.038$ & $-0.119\pm0.027$ & $-0.182\pm0.048$ & $-0.681\pm0.154$ & $-0.019\pm0.002$\\
 NGC 1340 & $  151$ & $ 10.99$ & $ 0.008\pm0.037$ & $-0.044\pm0.066$ & $-0.024\pm0.048$ & $-0.080\pm0.219$ & $-0.073\pm0.005$\\
 FCC  276 & $  280$ & $ 10.67$ & $-0.050\pm0.027$ & $-0.029\pm0.032$ & $-0.176\pm0.037$ & $-0.602\pm0.118$ & $-0.067\pm0.003$\\
 FCC  147 & $  264$ & $ 10.69$ & $-0.047\pm0.022$ & $ 0.025\pm0.039$ & $-0.154\pm0.029$ & $-0.516\pm0.090$ & $-0.042\pm0.003$\\
 IC  2006 & $   97$ & $ 10.38$ & $-0.027\pm0.048$ & $-0.173\pm0.060$ & $-0.171\pm0.074$ & $-0.515\pm0.232$ & $-0.123\pm0.004$\\
 FCC   83 & $  217$ & $ 10.51$ & $-0.002\pm0.021$ & $-0.108\pm0.031$ & $-0.089\pm0.037$ & $-0.248\pm0.115$ & $-0.105\pm0.002$\\
 FCC  184 & $  230$ & $ 10.82$ & $-0.037\pm0.046$ & $-0.089\pm0.048$ & $-0.013\pm0.079$ & $-0.074\pm0.248$ & $-0.011\pm0.001$\\
 FCC   63 & $  163$ & $ 10.43$ & $-0.045\pm0.035$ & $-0.082\pm0.040$ & $-0.171\pm0.046$ & $-0.513\pm0.167$ & $-0.120\pm0.006$\\
 FCC  193 & $   25$ & $ 10.48$ &          \nodata &          \nodata & $ 0.023\pm0.140$ & $ 0.056\pm0.475$ & $-0.107\pm0.003$\\
 FCC  170 & $   44$ & $ 10.29$ &          \nodata &          \nodata & $-0.137\pm0.074$ & $-0.578\pm0.350$ & $-0.007\pm0.005$\\
 FCC  153 & $   33$ & $  9.94$ &          \nodata &          \nodata & $-0.112\pm0.059$ & $-0.502\pm0.340$ & $ 0.078\pm0.012$\\
 FCC  177 & $   45$ & $  9.82$ &          \nodata &          \nodata & $-0.097\pm0.050$ & $-0.401\pm0.291$ & $ 0.200\pm0.014$\\
 FCC   47 & $  206$ & $  9.97$ & $-0.043\pm0.021$ & $-0.074\pm0.031$ & $-0.208\pm0.033$ & $-0.687\pm0.116$ & $-0.093\pm0.004$\\
 FCC  190 & $  106$ & $  9.87$ & $-0.087\pm0.022$ & $ 0.023\pm0.064$ & $-0.116\pm0.030$ & $-0.619\pm0.156$ & $-0.005\pm0.004$\\
 FCC  249 & $  115$ & $  9.99$ &          \nodata &          \nodata & $-0.115\pm0.041$ & $-0.446\pm0.165$ & $-0.083\pm0.007$\\
 FCC  148 & $   58$ & $ 10.03$ & $ 0.006\pm0.043$ & $ 0.019\pm0.049$ & $ 0.042\pm0.076$ & $ 0.161\pm0.305$ & $ 0.189\pm0.010$\\
 FCC  255 & $   53$ & $  9.56$ & $-0.035\pm0.034$ & $-0.044\pm0.041$ & $-0.147\pm0.044$ & $-0.526\pm0.179$ & $ 0.012\pm0.005$\\
 FCC  277 & $   22$ & $  9.88$ &          \nodata &          \nodata & $-0.300\pm0.119$ & $-0.995\pm0.501$ & $-0.055\pm0.003$\\
 FCC  182 & $   30$ & $  9.40$ &          \nodata &          \nodata & $ 0.080\pm0.078$ & $ 0.204\pm0.278$ & $-0.036\pm0.010$\\
\enddata
 \tablenotetext{1}{The name of galaxies}
 \tablenotetext{2}{Total number of GCs those $p>0.95$}
 \tablenotetext{3}{Logarithm of stellar mass (in unit of $\msun$)}
 \tablenotetext{4}{Color gradient of red GCs with error, if bimodal}
 \tablenotetext{5}{Color gradient of blue GCs with error, if bimodal}
 \tablenotetext{6}{Color gradient of all GCs with error}
 \tablenotetext{7}{Metallicity gradient of all GCs with error}
 \tablenotetext{8}{Color gradient of galaxy with error}
 \tablenotetext{Note}{Due to the FOV of ACS, the outer boundaries
                      of our measurements of gradients in most
                      galaxies are about 2--3 arcmin.}
\end{deluxetable*}
}


\begin{figure*}[!b]
  \begin{center}
  \includegraphics[width=0.7\textwidth]{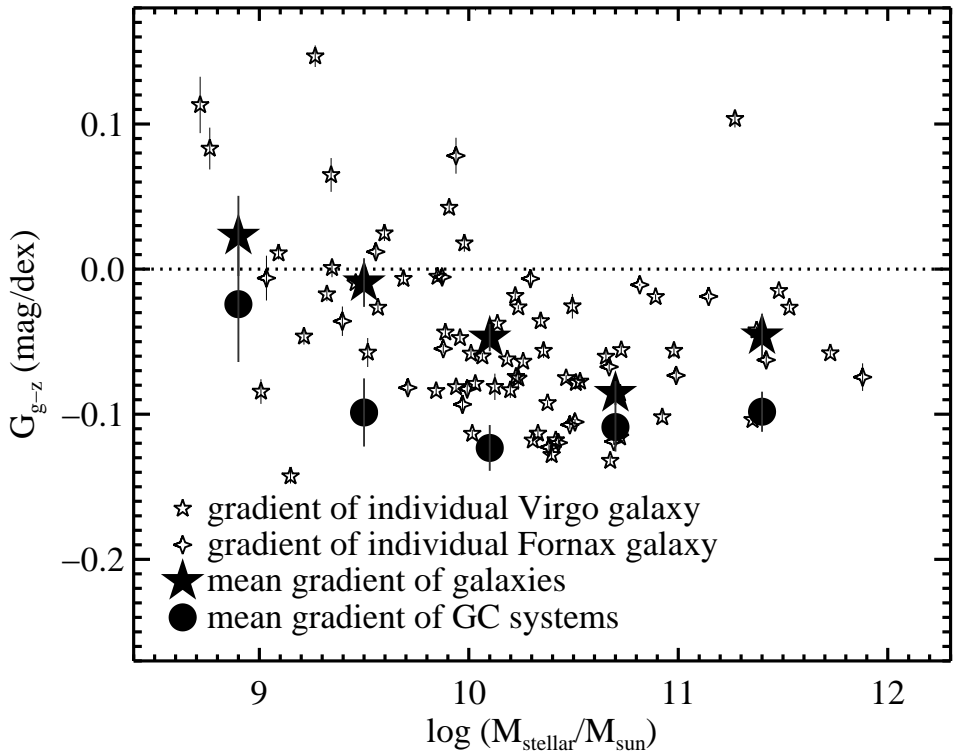}\\
  \caption{Color gradients of galaxies and GC systems binned by mass as a
    function of stellar mass, $\mstellar$. Small open stars
    with error bars are the gradients and errors in individual galaxies.
    Filled stars and circles describe the mean color
    gradients and errors of galaxies and GC systems in given mass bins.}
  \label{fig:compare}
  \end{center}
\end{figure*}

\subsection{The Gradient--Mass Relation and Comparison to Field Stars}

We have seen that the strength of the GC systems color gradient
varies as a function of galaxy mass.  Similar behavior has been seen
in the color gradients of the field stars for galaxies in other
studies.  One advantage of the ACSVCS and ACSFCS data is that we can
also measure color gradients for the host galaxies using the exact
same filters that we use for the GCs, allowing us to make a direct
comparison between the two galactic components.

\citet{Ferrarese2006_ApJS_164_334} measured the isophotal light
profiles of 100 early-type galaxies in the ACSVCS in both the $g$ and
$z$ band. The light profiles of 43 ACSFCS galaxies are measured by
using the same method (Ferrarese et al., in prep). The calculations
of the color gradients of these galaxies are based on their surface
photometry. To remove the effect of nuclei of early-type galaxies
(about $2\%$ of effective radius, see \citealt{Cot'e2006_ApJS_165_57,
Cot'e2007_ApJ_671_1456}) and eliminate the significant contamination
of sky background in outskirts, we measure the color gradients in the
range from $0.02~R_e$ to $R_e$. The definition of color gradient is
the same as that for GC systems (Equation~\ref{eqn:cgrad}). We list
the stellar color gradients for the ACSVCS/FCS galaxies in
Table~\ref{tab:cg_VCC}.

Figure~\ref{fig:compare} shows the relationship between the field
star color gradients and galaxy stellar mass, $\mstellar$. Larger
stars denote the mean gradient in a given mass bin. Similar to what
has been found in other studies, the color gradients of the
ACSVCS/FCS early-type galaxies are mostly negative, with low
mass galaxies having flat or positive gradients. We
find no significant difference between color
gradients of galaxies in the Virgo and Fornax.  In the mean, the
steepest gradients are found in intermediate-mass galaxies, which is
consistent with the finding of \citet{Tortora2010_MNRAS_407_144} from
SDSS surface photometry of galaxies. The circles in
Figure~\ref{fig:compare} show the mean values for color gradients of
GC systems, shown in bottom panel of Figure~\ref{fig:cg_mass}. We can
see that the color gradients of GC systems are systematically steeper
than those of the field stars, but that the gradient--mass relation
is similar in shape to that of stellar systems of galaxies.

At the bright end, the color gradients of GC systems seems to be
getting shallower again, but there is an important caveat. The ACS
observations of our sample galaxies have field view of
$3\farcm4\times3\farcm4$, which at the distance of the Virgo Cluster
is roughly 16~kpc on a side.  For the massive galaxies, we are only
probing the innermost regions of the halo, even when using fields
that observed neighboring galaxies. \citet{Rhode2001_AJ_121_210}, in
a wide-field study of the GC system in M49 (NGC~4472, VCC~1226),
found color gradients within $8\arcmin$, but also found that the
gradient disappeared when expanding the radius to $22\arcmin$.
\citet{Harris2009_ApJ_703_939} measured color gradients for the GCs
in M87 and found detectable gradients out to $\sim60$~kpc, or
$12\farcm5$, a roughly similar radial scale.  In this work, we have
mostly only calculated color gradients of the central part of the GC
systems of massive galaxies due to the limited field of view of ACS,
so for massive galaxies these color gradients are best described as
those for ``inner halo'' GCs.  A more comprehensive study of GC
system color gradients will require wide-field imaging, such as that
being taken for the Next Generation Virgo Cluster Survey (Ferrarese
\etal 2011, in prep.).

\section{Discussion}
\subsection{A Note on Projection Effects}

When we calculate the GC system color gradients, we use projected
galactocentric distances, not the true three-dimensional distances to
the galaxy centers. Projecting the GC system onto the plane of the
sky weakens the measured gradients, as GCs projected onto the center
are actually a mix of GCs at all radii.  This is less of a problem
for more centrally concentrated systems (i.e., more steeply rising
density profiles toward the center).  Because the distribution of red
GCs can be more concentrated than blue GCs, the projection effects
for the two subpopulations could be different.

\begin{figure}
  \includegraphics[width=0.48\textwidth]{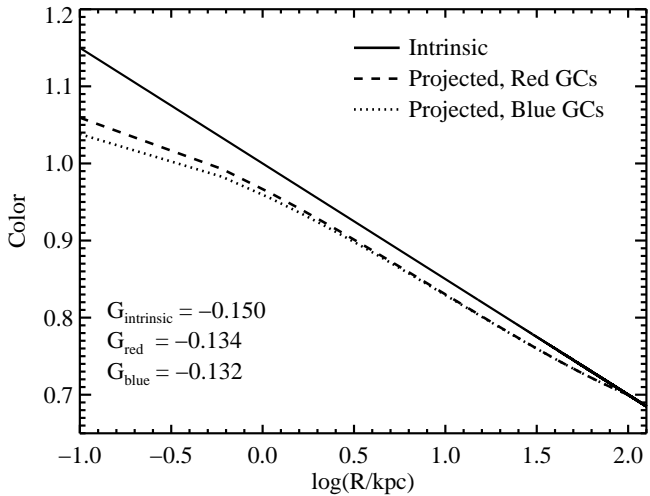}\\
  \caption{A simulated GC system color profile. Solid line is the intrinsic
  profile, dashed and dotted lines denote color profiles of red and
  blue GC populations when projected onto the plane of the sky.}
  \label{fig:model}
\end{figure}

In order to test the affects of projecting real gradients into the
plane of the sky, we provide one test case.
\citet{Cot'e2001_ApJ_559_828} deprojected the spatial distribution of
the red and blue GC subpopulations in the galaxy M87. The projections
of the model density distribution are consistent with the observed
surface density of GCs. They obtained:
\begin{eqnarray}
   n_{red}(r)  &=& (\frac{r}{3.3~{\rm kpc}})^{-1}(1+\frac{r}{3.3~{\rm kpc}})^{-2},~r<95~{\rm kpc} \\
   n_{blue}(r)  &=& (\frac{r}{20.5{\rm kpc}})^{-1}(1+\frac{r}{20.5{\rm kpc}})^{-2},r<125~{\rm kpc}
\end{eqnarray}
We assume that the initial color gradient of GC systems are $-0.15$
and project the model density distribution into two dimensions. The
resulting projected color profiles are shown in Figure
\ref{fig:model}. The projection effect flattens the gradient of both
red and blue GC systems to $-0.134$ and $-0.132$, respectively.
Because the blue GCs are more extended than the red ones, the
flattening is slightly more obvious in the gradient of blue GC
system. The total effect of projection in this case, however, is
relatively small, roughly 12\%, and the relative effect between the
red and blue GCs is negligible.  We do not correct for projection
effects because we do not know the three-dimensional density profiles
of the GC systems. We simply note that the true radial color
gradients for these GC systems are slightly steeper than the
projected gradients, but that for a realistic density profile of GCs,
this correction is likely to be of the order of $\sim10\%$.

\subsection{Gradient-Induced Bias in the Colors of GC Subpopulations}

Our results show that the GC systems of early-type galaxies display
significant negative color gradients, which is consistent with
previous work \citep[e.g.,][]{Strom1981_ApJ_245_416,
Geisler1996_AJ_111_1529, Rhode2004_AJ_127_302, Jord'an2004_AJ_127_24,
Cantiello2007_ApJ_668_209}. However, there exists more uncertainty about
whether individual GC subpopulations display color gradients or
not. Some previous studies have found shallow gradients in individual GC
subpopulations \citep[e.g.,][]{Bassino2006_A+A_451_789,
Harris2009_ApJ_703_939, Harris2009_ApJ_699_254} while other studies
have not \citep[e.g.,][]{Forbes2004_MNRAS_355_608,
Cantiello2007_ApJ_668_209, Kundu2007_ApJ_660_109}. In this work, we
find that only a few of the individual GC subpopulations show
significant gradients ($>3\sigma$), but the overall trend is obvious
when measured over 39 galaxies, i.e., individual GC subpopulations
display negative gradients statistically. It is the first homogeneous
study of color gradients of GC systems in early-type galaxies, and
emphasizes the power of using a large, homogeneous sample of
galaxies.

One consequence of this result could be on the mean measured colors
of the subpopulations.  The studies with the highest precision
photometry and least contamination have often used HST imaging
\citep[e.g.,][]{Larsen2001_AJ_121_2974, Peng2006_ApJ_639_95} which
necessarily has a small field of view relative to the largest nearby
GC systems. Results from these studies have shown that the mean color
of the blue and red GC subpopulations is a function of galaxy mass or
luminosity, where more massive galaxies have redder GCs, in the mean.
This has generally been interpreted as a mass--metallicity relation
for GC {\it systems} (as opposed to for individual GCs).  The
correlation for metal-poor GCs, although weaker than that for the red
GCs, has drawn particular interest because it implies a connection
between the earliest forming GCs and the final halos in which they
reside \citep{Larsen2001_AJ_121_2974, Burgarella2001_AJ_121_2647,
Strader2004_AJ_127_3431, Peng2006_ApJ_639_95}.

Given the fixed and relatively small field of view for the
instruments, there is a significant aperture sampling effect that
varies across the studied range of galaxy mass.
\citet{Peng2008_ApJ_681_197} showed that for galaxies with $M_B>-18$,
the entire GC system typically fits within the ACS field of view.  At
higher luminosities, the ACS field will miss some fraction of the
outer regions.  This fraction can be as high as $\sim90\%$, in the
case of M87.  This bias toward the centers of galaxies would not
matter if the GC subpopulations did not possess color gradients.  We
have found, however, that they typically have gradients of 0.04--0.05
mag/dex in ($g$--$z$).  This results in a bias where the most massive
galaxies are sampled where the GCs are most red.  This would
particularly affect the blue GCs, which can have a more extended
spatial distribution.

Although the resolution of this issue will
ultimately require precision wide-field imaging, we can estimate
the degree of bias that color gradients may have introduced.  For the
most massive galaxies in Virgo, such as M49, M87, and M60, the
effective radii ($R_{e, \rm GCs}$) of the GC populations are 42, 41,
and 24~kpc, respectively. This was determined using \sersic\ fits to
the GC spatial distribution from ACSVCS data and published photometry
\citep{McLaughlin1999_AJ_117_2398, Rhode2001_AJ_121_210}. The mean
projected radius for the GCs observed in the ACS field, and for which
the subpopulations colors were measured in
\citet{Peng2006_ApJ_639_95}, is roughly 5~kpc. Given the gradients
for the red and blue populations measured in this paper
(Table~\ref{tab:cg_VCC}), we can infer the expected difference
between the mean color in the ACS field and the mean color at $1R_{e,
\rm GCs}$, which should roughly represent the mean color of the
entire GC subpopulation {\it if the color gradient is constant at all
radii}. The color difference, $\Delta_{g-z}(ACS-R_{e, \rm GCs})$, for
M49, M87, and M60 is: $0.031$~mag, $0.046$~mag, and $0.019$~mag,
respectively.

Could such a shift to the blue affect the previously published
correlations between galaxy mass and GC subpopulation metallicity? We
estimate $\Delta_{g-z}$ as a function of galaxy $M_B$ in the ACSVCS
sample for easy comparison with the analysis in
\citet{Peng2006_ApJ_639_95}.  We use the measured mean colors within
the HST/ACS field of view from Table~4 of
\citet{Peng2006_ApJ_639_95}, $R_{e, \rm GCs}$ for the GC systems in
the ACSVCS galaxies (\citealt{Peng2008_ApJ_681_197}; Peng et al., in
prep), and the mean color gradients for the red and blue GC
subpopulations ($-0.048$ and $-0.041$, respectively,
Figure~\ref{fig:cg_mass}) to infer the mean color for the red and
blue subpopulations at $R_{e, GCs}$, which we take to be
representative of the entire population. We weight each galaxy's
contribution to $\Delta_{g-z}$ by their total number of GCs from
\citet{Peng2008_ApJ_681_197}.

\begin{figure}[]
  \includegraphics[width=0.48\textwidth]{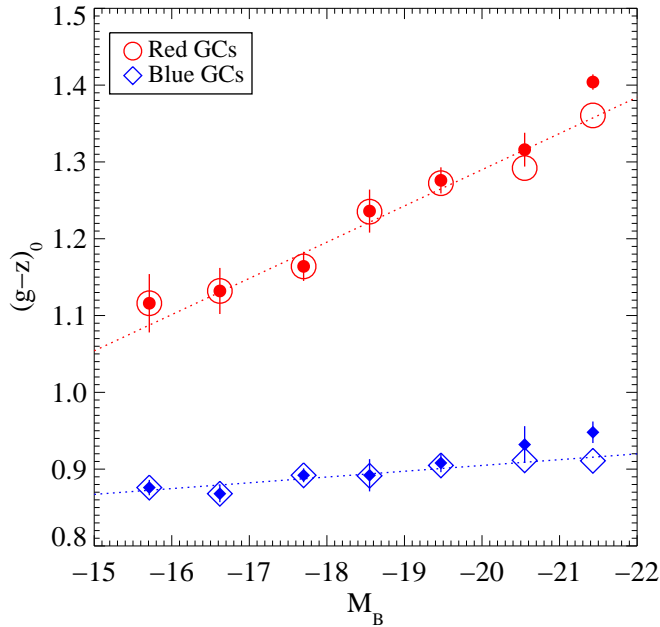}\\
  \caption{Mean colors of red (circles) and blue GCs  (diamonds) as a
    function of galaxy
    luminosity ($M_B$) with data from Peng \etal (2006) (filled
    points) and inferred
    corrections to a common radius of $R_{e, \rm GCs}$ (open points).  The
    correction is only important for
    the most luminous galaxies.  The dotted lines show linear fits to
    the data with inferred correction.  The slope of the blue GCs
    is -0.007, only half the value when fitting the original
    data.}\label{fig:cg_corr}
\end{figure}

Figure~\ref{fig:cg_corr} mirrors Figure~8a of
\citet{Peng2006_ApJ_639_95} and plots the mean colors versus galaxy
$M_B$, both measured with ACS (solid points) and inferred at $R_{e,
\rm GCs}$ (open points).  As expected, the correction is only
important for the two most luminous bins in $M_B$. Whereas the
relation for the red GCs is not significantly affected, as it was
originally quite steep, the slope for the blue GC relation is
noticeably smaller.  The fit to the original measurements gave a blue
GC slope of $-0.0126\pm0.0025$, a nonzero slope at the $5\sigma$
level. The fit to the newly inferred mean colors produces a slope of
$-0.0069\pm0.0025$ (systematic errors from the correction process are
not included).  This shallower slope is now only significant at
$2.8\sigma$, and could potentially be even less significant if more
accurate color gradients on the outer regions are measured to be
steeper than what we have measured.

We want to emphasize that this exercise is far from conclusive, and
only serves to warn that color gradients in the GC subpopulations
will potentially affect conclusions drawn from imaging the central
regions of galaxies.  Similar biases have been noted for the
color-magnitude relation of early-type galaxies where colors are
measured in fixed apertures \citep{Scodeggio2001_AJ_121_2413}.  The
correction that we infer out to $R_{e, \rm GCs}$ assumes a constant
color gradient over the entire GC system. This assumption is
unverified, and probably provides an upper limit on the possible
correction, given the results of \citet{Rhode2001_AJ_121_210} and
\citet{Harris2009_ApJ_699_254} who find a flattening gradient at
large radius. Nevertheless, a shallower (or potentially non-existent)
relation between the mean color (metallicity) of metal-poor GCs and
their host galaxy mass will have implications for understanding the
formation of GC systems, and the solution to this problem awaits
high-quality wide-field data (e.g., \citealt{Rhode2004_AJ_127_302}).

\subsection{The Formation of GC systems and Their Hosts}
That GC systems should have negative color (metallicity) gradients is
perhaps not surprising given that GC formation is by its very nature
a product of high star formation efficiency.  Most models predict
that high efficiency of star formation plus metal retention leads to
more enriched populations at the centers of galaxies.  Interestingly,
even though GCs are among the oldest objects in galaxies, and thus
have presumably experienced the largest amount of merger-induced
radial mixing of any stellar population, the color gradients in most
intermediate- and high-mass galaxies are still significantly
negative.

\citet{DiMatteo2009_A+A_499_427} investigated the survival of
metallicity gradient after a major dry merger. For ellipticals with
similar initial gradients, they concluded that the final gradient is
about 0.6 times of the initial after a major dry merger.
Dissipational mergers, however, can either flatten or enhance
gradients, depending on the initial gradients and the amount of gas
involved. This dependence on merger history is one of the reasons why
gradients in massive galaxies are expected to have larger dispersion.
We have very few galaxies on the high-mass end, so it is difficult
for us to probe the dispersion in GC system color gradients in this
mass range. For the high-mass galaxies in our sample, we are also
only probing the very inner halo, so it is possible that the
gradients in this region are either more robust to dilution or had
the strongest initial gradients.  The color gradients for blue GCs
(presumably the oldest GCs) in massive galaxies are detectable but
shallower than at lower masses, and this may be a sign that mergers
have played a part in their history.  It would be interesting to
extend this study to wider fields of view for the more massive
galaxies.

We find that the mean gradients for the red and blue subpopulations
are similar in magnitude ($-0.048$ and $-0.041$, respectively), but
its interpretation is confounded by the uncertain conversion from
($g-z$) to metallicity.  Both the \citet{Peng2006_ApJ_639_95} and
\citet{Blakeslee2010_ApJ_710_51} transformations have slopes that are
roughly 3--4 times steeper at the mean blue GC color than at the mean
red GC color.  By extension, the true metallicity gradient for
metal-poor GCs should be 3--4 times steeper than that for metal-rich
GCs (given their similar gradients in color).  This would be a fairly
remarkable difference between the two populations, but is still
entirely dependent on the assumed color-metallicity relation.  We
plan to revisit this question when the transformation from ($g$--$z$)
is better understood.

The relationship between the color gradients of GC systems and host
galaxy mass offers some interesting insights into the formation and
evolution of stellar halos in early-type galaxies.  Even for
measurements of color gradients from galaxy surface photometry, it
was only relatively recently that the data quality and mass range
probed has been sufficient to study trends in galaxy mass (e.g.,
\citealt{Forbes2005_MNRAS_361_6, Spolaor2009_ApJ_691_138,
Rawle2010_MNRAS_401_852, Tortora2010_MNRAS_407_144}).
\citet{1009.2500} have also run simulations to show that environment
can also play a role in the observed gradients. Our results show that
the shape of gradient--mass relation for GC systems is similar to
that for the galaxies themselves, with a minimum around
$\approx10^{10}\msun$ (Figure~\ref{fig:cg_mass}). That the GC color
gradients are universally steeper than those for the field stars is
an interesting result.  If the GCs formed in higher efficiency star
forming events than the bulk of the field stars (e.g.,
\citealt{Peng2008_ApJ_681_197}), then that might result in steeper
gradients.  One caveat for the interpretation, however, is that the
total GC gradients are actually a combination of the red and blue GC
populations, which may not have formed contemporaneously in the
present-day halo.  The steeper gradients are likely a combination of
the increasing fraction of blue GCs and the increasing specific
frequency of GCs at lower metallicity \citep{Harris2002_AJ_123_3108}.
The color gradients for the individual GC populations are similar in
magnitude, if not slightly shallower than the gradients for the field
stars.

The shape of the gradient--mass relation for both GC systems and
field stars is broadly consistent with a model where color
(metallicity) gradients are increasingly steeper in higher mass halos
due to metal retention, but then are diluted in the highest-mass
galaxies ($\mstellar\gtrsim 10^{10}\msun$) due to the increasing
importance of mergers in their evolution. One difference between the
GCs and the stars is that the stars in some dwarfs exhibit
significantly positive color gradients, which are often interpreted
as due to age gradients (age increasing with radius)
\citep[e.g.,][]{LaBarbera2009_ApJ_699_76,
Spolaor2010_MNRAS_408_272}. This is not difficult to produce if
there has been recent low level star formation at the galaxy center.
However, we do not see any case of this for the GC systems, nor might
we expect to as the star formation rate density required to produce
young GCs is much higher than needed to produce a slight age gradient
in the field. We notice that there is a prominent outlier in
Figure \ref{fig:compare}, VCC 798 (M85) with mass
$\sim10^{11.27}\msun$, which has a steep positive color gradient.
This galaxy is known as a young, gas-rich merger remnant
\citep[e.g.,][]{Schweizer1992_AJ_104_1039, Peng2006_ApJ_639_838} and
hosts a large-scale stellar disk \citep{Ferrarese2006_ApJS_164_334}.
During the gas-rich merger, the central starburst produced a young,
blue stellar population in the center of galaxy. Therefore, the
positive color gradients are common in gas-rich merger remnant
\citep[e.g.,][]{Yamauchi2005_MNRAS_359_1557}. However, the color
gradient of the GC system in VCC 798 is negative and quite normal. One of the
possible reasons is that the number of GCs formed during the merger is negligible
compared to the preexisting old GC population. The
fact that the GC systems of dwarf galaxies have shallow or flat color
gradients suggests that metal retention and mixing were not efficient
during the epoch of GC formation.

\section{Conclusion}
\label{sec:conclusion}

We use HST imaging from the ACS Virgo and Fornax Cluster Surveys to
conduct the first large-scale study of globular cluster system radial
color gradients.  We present results for 76 early-type galaxies,
measuring ($g$--$z$) color gradients for GC systems across a range in
galaxy stellar mass ($8.7<\log(\mstellar / \msun)<11.8$).  For 39
galaxies whose GC systems show significantly bimodal color
distributions, we also measure the color gradients in the GC
subpopulations.  Using the surface photometry of ACSVCS galaxies from
\citet{Ferrarese2006_ApJS_164_334}, we measure the radial color
gradients of the field stars in the same galaxies and same filters,
allowing a direct comparison of GC and field star radial gradients.
We caution that the FOV of ACS means we only measure the central part
of many large galaxies, which may introduce an aperture bias if the
color gradient of galaxies are not constant with the radius. We find
that:

\begin{enumerate}
  \item GC systems as a whole have negative color gradients, with
      an average gradient over the entire sample of
      $-0.112\pm0.009$~mag in ($g-z$) per dex in radius.

  \item On average, red and blue GC subpopulations also show
      significantly negative color gradients at the mean level of
      $-0.048\pm0.010$ and $-0.041\pm0.006$, respectively.
      Although a gradient is sometimes difficult to detect for
      any individual galaxy, the combined sample shows this
      property with higher signal-to-noise.

  \item We find a relationship between GC system gradient
      strength and galaxy stellar mass, where the gradients are
      flat at low mass, increasingly negative with mass until
      $\mstellar \approx10^{10}\msun$ and then staying constant
      or less negative at higher mass.  This trend parallels the
      gradient--mass relationship we find for the field stars in
      the ACSVCS galaxies. The GC system gradients are
      systematically steeper than that for the field stars, which
      is likely a reflection of the dominance of blue GCs at
      large radius. These observed trends, however, are limited
      by the small number of galaxies at high and low mass in our
      sample.

  \item Color gradients in the GC subpopulations can cause a bias
      in the measurement of the mean colors of GCs when the data
      only covers the central region of the galaxy.  We infer a
      correction using the measured gradients and find that the
      slope between the mean color of metal-poor GCs and the
      luminosity of their hosts can be reduced by nearly a factor
      of two from previous measurements, raising questions about
      its level of significance.

  \item The shape of the gradient--mass relation for GC systems
      is consistent with picture where the formation and chemical
      enrichment of the GC system becomes more efficient as the
      mass of the host galaxy increases, but is further affected
      by significant merging and radial mixing in the most
      massive galaxies.

  \item In a test case, the intrinsic, three-dimensional color
      gradients are likely to be roughly $\sim10\%$ steeper than
      the projected gradients given a reasonable spatial
      distribution of GCs.
\end{enumerate}

\acknowledgments

The authors thank the anonymous referee for useful comments.
C.~L.\ would like to thank the Peking University Postdoctoral Fund
for its support. E.~W.~P.\ acknowledges support from the Peking
University 985 Fund, and grant 10873001 from the National Natural Science
Foundation of China. A.~J.\ acknowledges support from Fondecyt
project 1095213, BASAL CATA PFB-06, FONDAP CFA 15010003 and MIDEPLAN
ICM Nucleus P07-021-F.

Support for programs GO-9401 and GO-10217 was provided through grants
from the Space Telescope Science Institute, which is operated by the
Association of Universities for Research in Astronomy, Inc., under
NASA contract NAS5-26555.

This publication makes use of data products from the Two Micron All
Sky Survey, which is a joint project of the University of
Massachusetts and the Infrared Processing and Analysis
Center/California Institute of Technology, funded by the National
Aeronautics and Space Administration and the National Science
Foundation.

This research has made use of the NASA/IPAC Extragalactic Database
(NED) which is operated by the Jet Propulsion Laboratory, California
Institute of Technology, under contract with the National Aeronautics
and Space Administration.

Facilities: \facility{HST(ACS)}


\clearpage
\end{document}